# MACHO COLLABORATION SEARCH FOR BARYONIC DARK MATTER VIA GRAVITATIONAL MICROLENSING

K. Griest[*,†], C. Alcock[∥,†], R.A. Allsman[∥], T.S. Axelrod[‡], D.P. Bennett[†,∥],
K.H. Cook[∥,†], K.C. Freeman[‡], J. Guern[*,†], M. Lehner[*,†], S.L. Marshall[†,♭],
S. Perlmutter[†], B.A. Peterson[‡], M.R. Pratt[†,♭], P.J. Quinn[‡],
A.W. Rodgers[‡], C.W. Stubbs[†,♣], W. Sutherland[♠], D. Welch[♡]
(The MACHO Collaboration)

[*] Department of Physics, University of California, San Diego, CA 92039
[†] Center for Particle Astrophysics, University of California, Berkeley, CA 94720
[∥] Lawrence Livermore National Laboratory, Livermore, CA 94550
[‡] Mt. Stromlo and Siding Spring Observatories,
Australian National University, Weston, ACT 2611, Australia
[♭] Department of Physics, University of California, Santa Barbara, CA 93106
[♣] Department of Astronomy, University of Washington, Seattle, WA 98195
[♠] Department of Physics, University of Oxford, Oxford OX1 3RH, U.K.
[♡] Department of Physics and Astronomy, McMaster University,
Hamilton, ON L8S 4M1, Canada

I present results from the MACHO collaboration gravitational microlensing search. I describe the experiment and the nearly 50 microlensing events that have been detected. Limits on the baryonic content of the halo are given, as are estimates of the Macho contribution to the dark halo. Optical depths toward the bulge, and several unusual events such as a binary lens and a parallax event are discussed. Possible interpretations of these results are also discussed.

## 1. Introduction

It is remarkable that science can discover exotic states of matter such as the top quark, and detect one part in $10^5$ fluctuations in the microwave background and still not know what the primary constituent of the Universe is. Over the past decades the existence of this mysterious dark matter has become accepted, as evidence for its gravitational dominance has become overwhelming. It is "seen" on scales from small dwarf galaxies, to large spiral galaxies like our own, to clusters of galaxies, to the largest scales yet observed. The best evidence for dark matter comes from the rotation curves of spiral galaxies like our own Milky Way. Using 21 cm emission to detect neutral hydrogen far beyond the visible edge of galaxies, it is clear that 3 to 10 times as much dark matter as visible matter must exist in these systems. Translating these results into $\Omega$, the density averaged over the Universe

---

[★] 

divided by the critical density, one finds $\Omega_{halos} > \sim 0.03$. On larger scales, such as in clusters or from large scale flows, larger amounts of dark matter are found (for example, $\Omega > 0.2$), but the results are less robust.[1,2] When compared with density of luminous matter, $0.003 < \Omega_{lum} < 0.008$, the extent of the dark matter problem becomes clear.

One would like to know what this dark material is, but how does one detect something that neither emits nor absorbs electromagnetic radiation at any known wavelength? If the dark matter consists of exotic elementary particles such as Wimps[3] or axions,[4] then the only way is via particle accelerators or special purpose laboratory experiments. However, if the dark matter consists of astronomical objects in the $10^{-8} M_\odot$ to $10^3 M_\odot$ range, then the 1986 idea of Paczyński[5] of using gravitational microlensing can be applied. Now, constraints from big bang nucleosynthesis imply that if $\Omega_{tot} = 1$, then most of the dark matter must be non-baryonic; that is, consist of exotic elementary particles. However, as seen from the numbers above, all the dark matter in the halos of spiral galaxies could be baryonic and still satisfy the nucleosynthesis constraint $0.01 < \Omega_{baryon} < 0.1$. Since all dark matter detection experiments, including the searches for Wimps and axions, search only for dark matter in the Milky Way halo, and since it is the spiral galaxy halo dark matter which is by far the most secure, it should be clear that a search for baryonic dark matter is essential. Candidate baryonic objects include brown dwarf stars, jupiter mass objects, or remnants from an early generation of star formation; basically any astronomical objects which would now be too small or too dim to be seen at large distances. The remarkable thing about microlensing experiments is that they are sensitive to all such objects over a huge range of masses.

There are now several groups, including our own, which are returning results from searches for baryonic dark matter, and it is these results which I want to discuss. While it is now clear that we have not discovered enough dark matter to make up a full standard dark halo, we seem to have discovered a substantial new component of our galaxy, and may well have found a part of the long-sought dark matter.

## 2. Microlensing

In order to explain the roughly flat rotation curve of the Milky Way, a large dark halo, usually spherical, extending to at least 50 kpc is hypothesized. The density of this halo should drop roughly as $r^{-2}$ at large radii to produce a flat rotation curve, and the objects making up this halo should move with velocities in the 270 km/sec range in order to satisfy hydrostatic equilibrium. This is true whether the objects are elementary particles or massive compact halo objects (Machos).

The idea of microlensing rests upon Einstein's observation that if a massive object lies directly on the line-of-sight to a much more distance star, the light from the star will be lensed and form a ring around the lens. This "Einstein ring" sets the scale for all the microlensing searches, and in the lens plane, the radius of that ring is given by

$$r_E = 610 R_\odot [\frac{m}{M_\odot} \frac{L}{\text{kpc}} x(1-x)]^{1/2},$$

where $R_\odot$ and $M_\odot$ are the solar radius and mass, $m$ is the Macho mass, $L$ is the distance to the star being monitored, and $x$ is the distance to the Macho divided by $L$.

In fact, it is extremely unlikely for a Macho to pass precisely on the line-of-sight, but if there is a near miss, two images of the star appear separated by a small angle. For masses in the stellar range and distances of galactic scale this angle is too small to be resolved, but the light from both images add and the star appears to brighten. The amount of brightening can be large, since it is roughly inversely proportional to the minimum impact parameter $b/r_E$. Since the Macho, Earth, and source star are all in relative motion, the star appears to brighten, reaches a peak brightness, and then fades back to its usual magnitude. The brightening as a function of time is called the "lightcurve" and is given by

$$A(u) = \frac{u^2 + 2}{u\sqrt{u^2 + 4}}, \qquad u(t) = [u_{min}^2 + (2(t - t_0)/\hat{t})^2]^{1/2},$$

where $A$ is the magnification, $u = b/r_E$ is dimensionless impact parameter, $t_0$ is the time of peak amplification, $\hat{t} = 2r_E/v_\perp$ is the duration of the microlensing event, $v_\perp$ is the transverse speed of the Macho relative to the line-of-sight, and $u_{min}$ is value of $u$ when $A = A_{max}$. Thus the signature for a microlensing event is a time-symmetric brightening of a star occurring as a Macho passes close to the line-of-sight. When a microlensing event is detected, one fits the lightcurve and extracts $A_{max}$, $t_0$, and $\hat{t}$. The primary physical information comes from $\hat{t}$, which contains the Macho velocity, and through $r_E$ the Macho mass and distance. Unfortunately, one cannot uniquely find all three pieces of information from the measurement of $\hat{t}$. However, statistically, one can use information about the halo density and velocity distribution, along with the distribution of measured event durations to gain information about the Macho masses. Using a standard model of the dark halo, Machos of jupiter mass ($10^{-3} M_\odot$) typically last 3 days, while brown dwarf mass Machos ($0.1 M_\odot$) cause events which last about a month.

In order to perform the experiment, a large number of stars must be followed, since, assuming a halo made entirely of Machos, the probability of any Macho crossing in front of a star is about $10^{-6}$. Thus many millions of stars must be monitored in order to see a handful of microlensing events. In addition, if one wants to see microlensing from objects in the dark halo, the monitored stars must be far enough away so that there is a lot of halo material between us and the stars. Therefore, the best stars to monitor are those in the Large and Small Magellanic Clouds (LMC and SMC) at distances of 50 kpc, and 60 kpc respectively, and also stars in the galactic bulge, at 8.5 kpc.

There are several experimental groups that have undertaken the search for microlensing in the LMC and galactic bulge and have returned results. The EROS collaboration, has reported 3 events towards the LMC[6], the OGLE group has reported about 15 events towards the bulge,[7] and the DUO collaboration has about a dozen preliminary events towards the bulge.[8] Our collaboration has seen about 5 events towards the LMC,[9,10,11] and about 50 events towards the bulge.[12,13,14] We are also monitoring the SMC, but have yet to analyze that data. In what follows I will concentrate on MACHO collaboration data.

## 3. MACHO Experiment

The MACHO project has full-time use of the 1.27-meter telescope at Mount Stromlo Observatory, Australia, for a period of at least 4 years from July 1992. In order to maximize throughput a dichroic beamsplitter and filters provide simultaneous images in two passbands, a 'red' band (approx. 5900–7800 Å) and a 'blue' band (approx. 4500–5900 Å). Two very large CCD cameras[15] are employed at the two foci; each contains a 2 × 2 mosaic of 2048 × 2048 pixel Loral CCD imagers, giving us a sky coverage of 0.5 square degrees. Observations are obtained during all clear nights and part nights, except for occasional gaps for telescope maintenance. The default exposure times are 300 seconds for LMC images, 600 sec for the SMC and 150 seconds for the bulge, so over 60 exposures are taken per clear night. As of 1995 April, over 30000 exposures have been taken with the system, of which about 19000 are of the LMC, 2000 of the SMC and 9000 of the bulge. The images are taken at standard sky positions, of which we have defined 82 in the LMC, 21 in the SMC and 75 in the bulge.

Photometric measurements from these images are made with a special-purpose code known as SoDoPHOT,[16] derived from DoPHOT[17]. For each star, the estimated magnitude and error are determined, along with 6 other parameters (quality flags) measuring, for example, the crowding, and the $\chi^2$ of the point-spread-function fit. It takes about an hour on a Sparc-10 to process a field with 500,000 stars, and so with the computer equipment available to us we manage to keep up. The set of photometric datapoints for each field are re-arranged into a time-series for each star, combined with other relevant information including the seeing and sky brightness, and then passed to an automated analysis to search for variable stars and microlensing candidates. The total amount of data collected to date is more than two Terabytes, but the time-series database used for analysis is only about 100 Gbytes.

## 4. Event Selection

Most of the stars we monitor are constant within our photometric errors, but about one half of one percent are variable. The MACHO database, as repository for the largest survey ever undertaken in the time domain, is an extremely valuable resource for studies of variable stars. From our first year LMC data alone we have already identified about 1500 Cepheid variables, 8000 RR Lyrae, 2200 eclipsing binaries, and 19000 long period variables. Example lightcurves from each of these classes are shown in Figure 1. We also have many rare variables, and have given the first conclusive evidence of 1st overtone pulsation in classical Cepheid's[18] We have also observed what may turn out to be entirely new types of variable stars.[19]

Given that the incidence of stellar variability, systematic error, and other sources of stellar brightening is much higher than the incidence of microlensing, how can one hope to discriminate the signal from the background among the tens of millions of stars we monitor nightly? Fortunately, there are several very powerful microlensing signatures which exist:

1. High amplification. Very high amplifications are possible, so we can set our

**Figure 1.** Example phased variable star lightcurves from first year LMC data. The top row shows RR Lyrae variables of type RRe, RRc, and RRab. The second row shows a range of periods for eclipsing binaries. The third row shows an overtone and a fundamental mode Cepheid and a low amplitude Long Period Variable.

$A_{max}$ threshold high enough to avoid many types of systematic error background.

2.) Unique shape of lightcurve. Only 5 parameters are needed to completely specify the 2-color lightcurve.

3.) Achromaticity. Lensing magnification should be equal at all wavelengths, unlike brightenings caused by most types of stellar variability.

4.) Microlensing is rare. The chance of two microlensing events occurring on the same star is so small, that any star with more than one "event" can be rejected as a microlensing candidate.

5.) Statistical tests: The distribution of peak magnifications $A_{max}$ is known a priori. Microlensing should occur with equal liklihood on every type and luminosity of star, unlike known types of stellar variability. New microlensing events should be discovered each year at a constant rate.

6.) Alert possibility. Our alert system is now working and we can catch microlensing before the peak and get many measurements of high accuracy. Other spectral and achromaticity tests can also be performed in follow-up mode.

Using these criteria, as well as others, we have found it possible to pick out microlensing candidates from variable stars, etc. For example, starting with about 9.5 million lightcurves from our first year LMC database, we remove all but 40000 by requiring a significant bump in both the red and blue lightcurves, and removing the reddest 0.5% of stars (which are mostly long period variables). Then performing a microlensing fit, and making cuts on statistics such as the $\chi^2$ per degree of freedom of the fit, the "crowding" of the star, the number of points in the peak, and the duration of the event, etc. we reduce the number of candidates down to a few thousand. Our last two cuts are made on the maximum magnification and the improvement in fit $\chi^2$ when a microlensing fit is compared to a fit to a constant star. We are left with a handful of events. Several of these are nebulosity junk, long time scale bumps on faint stars in the middle of high nebulosity regions. For our efficiency calculation, an automated way of removing such events was developed. Two of the remaining events are in fact the same star which happened to reside in a field overlap. So we are left with three microlensing candidates. These are shown in Figure 2.

**Figure 2:** The three observed stellar lightcurves that we interpret as gravitational microlensing events are each shown in relative flux units (red and blue) vs time in days. The solid lines are fits to the theoretical microlensing shape.

One of these events is clearly superior in signal/noise to the others, and we have confidence in the microlensing label. It has $A_{max} = 7.2$, and $\hat{t} = 35$ days. The other two, while passing all our cuts, and certainly consistent with microlensing, are less certain to be actual microlensing. We should note that our alert system has found a couple more high signal/noise LMC microlensing events, which are not included here, since we have performed efficiency calculations only on the first year data set.

**Figure 3.** Example lightcurves from first year bulge data. Four of the 43 microlensing events are shown.

Now, if we had found only these 3 events towards the LMC, we would not be as confident as we are, that we have seen microlensing. However, we have many more events towards the galactic bulge, and some of these are of incredibly high signal/noise. We cannot use the same selection criteria for the bulge as for the LMC since our observing schedule towards the bulge is different, and the bulge stellar population, distance, crowding, and extinction are different, but using the same statistics, we can make a similar selection procedure. We find about 43 candidates in our first year data (and since then a dozen or so more in our alert system). Examples of lightcurves from the bulge are shown in Figure 3. Some of these events are truly beautiful, with durations of many months and magifications of almost 20. Coupled with the dozen events from the OGLE collaboration, I think little doubt remains that microlensing has been seen.

## 5. Efficiency

What do the microlensing events mean for the dark matter question? In order to answer, we need to know the efficiency with which our system can detect microlensing. This is a non-trivial calculation. In order to calculate our efficiencies, we add simulated stars to real images, and then artificially brighten them. We run the photometry code on these simulated images and find what the photometry code returns when a star brightens under different atmospheric and crowding conditions. These results are incorporated into a large Monte Carlo in which simulated microlensing events are added to our actual (non-microlensing) data and fed into the same time-series analysis and selection procedure which produced the 3 LMC microlensing candidates. Thus we have explicitly taken into account inefficiencies caused by bad weather and system down time, our analysis and selection procedure, as well as blending of the underlying stars due to crowding, and systematic errors in our photometry. Since in order to calculate the expected number of events, we need to need to integrate a theoretical microlensing rate over our measured efficiency, we need efficiency $\epsilon$, as a function of $\hat{t}$. This is shown in Figure 4.

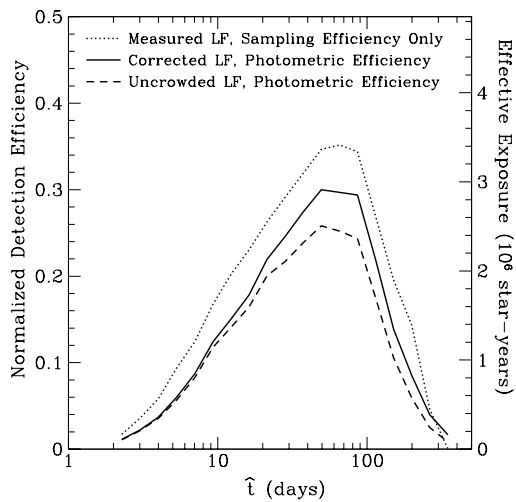

**Figure 4.** Microlensing detection efficiency from first year LMC data. The "corrected" solid curve is the best estimate. From reference 10.

The number of expected events is

$$N_{exp} = E \int \frac{d\Gamma}{d\hat{t}} \epsilon(\hat{t}) d\hat{t},$$

where our total exposure $E_{LMC} = 9.7 \times 10^6$ star-years, and $d\Gamma/d\hat{t}$ is a differential microlensing rate which can be calculated given a model of the dark halo.[20,21] We see that our efficiencies peak at about 30% for month long events. For short events, we do not have enough data points to make a sure microlensing identification, and our selection criteria explicitly exclude long duration events, since we need a good baseline to differentiate lensing events from variable stars. We should note that while these efficiencies seem low, we have checked by eye, and very few "good looking" microlensing candidates escape our selection criteria. There are substantial gaps in our data, and many stars are not measured well enough due to crowding or faintness to provide a good background source.

## 6. Interpretation of LMC Events

Using our sample of microlensing events, there are two complementary analyses which can be performed. First, we can set a conservative limit on the Macho contribution to the dark halo. Since we know our efficiencies, and we have certainly not seen more than 3 microlensing events from halo objects, any halo model which predicts more than 7.75 events can be ruled out at the 95% C.L. This result will be independent of whether or not all three candidate events are due to microlensing, and independent of whether or not the lenses are in the dark halo. Second, if we make the further assumption that all three events are due to microlensing of halo objects, we can estimate the mass of the Machos and their contribution to the mass of the dark halo.

In order to do either analysis we need a model of the dark halo. We need to know the total mass of the halo, and we need the density and velocity distribution to calculate an expected microlensing rate. The main constraints on the halo come from the Milky Way rotation curve, which is not as well determined as rotation curves in other galaxies. Constraints from the orbits of satellite galaxies also exist, but there is considerable uncertainty in both the total halo mass and the expected microlensing rate coming from uncertainty in the size and shape of the Milky Way halo.[22,23,21] Using a very simple, but commonly used halo model,[20] we can calculate the number of expected events as described above, and the results are shown in Figure 5. If the Milky Way has a standard halo consisting entirely of Machos of mass $0.001 M_\odot$ then we should have seen more than 20 events, with fewer events at larger or smaller masses. However, even if the halo dark matter consists of Machos, it is very unlikely that they all have the same mass. Fortunately, it can be shown,[20] that if one rules out all halos made of unique Macho mass between masses $m_1$ and $m_2$, then one has ruled out a halo consisting of ANY distribution of masses as long as only masses between $m_1$ and $m_2$ are included. Thus we can make the powerful conclusion that a standard halo consisting of any objects with masses between $8 \times 10^{-5} M_\odot$ and $0.3 M_\odot$ has been ruled out by our first year data.

As mentioned above, there is no strong reason to believe that the Milky Way halo is precisely as specified in the standard halo, and we would like to test the robustness of the important results above by considering a wider range of viable halo models. To this end, we have investigated a class of halo models due to Evans.[24] These models have velocity distributions which are consistent with their density profiles, and allow for halos which are non-spherical, and which have rotation curves which gently rise or fall. A description of the parameters that specify these models, along with microlensing formulas can be found in Alcock, *et al.*.[21] Basically we consider models which give rotation velocities within 15% of the IAU standard value of 220 km/sec, at the solar circle (8.5 kpc) and twice the solar circle. As pointed out by Evans and Jijina,[25] the contribution of the stellar disk plays an important role in the predicted microlensing rate. This is because much (or even most) of the rotation speed could be due to material in the disk, so we consider various size disks, as well. Table 1 shows parameters for a set of 8 disk/halo models, and Table 2 shows the resulting microlensing predictions if such a halo consisted entirely of Machos. Model "S" is the standard halo, while models A through G are Evans models.

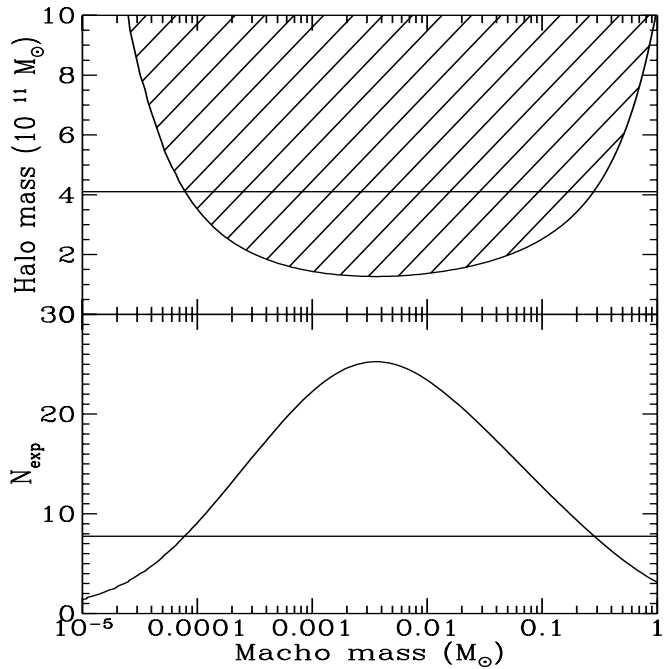

**Figure 5:** The lower panel shows the number of expected events predicted from the standard model halo with a delta function mass distribution. Given three observed events, points above the line drawn at $N_{exp} = 7.7$ are excluded at the 95% CL. The upper panel shows the 95% CL limit on the halo mass in MACHOs within 50 kpc of the galactic center for the model. Points above the curve are excluded at 95% CL while the line at $4.1 \times 10^{11} M_\odot$ shows the total mass in this model within 50 kpc. From Reference 10.

In Figure 6, we show the expected number of events in the above models as a function of Macho mass. Strong limits are found on heavy halo models such as B, while only very weak limits are found on very light halo models such as C and E, F, and G. This is because microlensing is sensitive not to the total mass in the halo, but only to the mass in Machos. So one can get a much more model independent limit on the Macho content of the halo by limiting the *total mass in Machos*, rather than the *Macho fraction* of the halo. This is shown in the upper panel of Figure 5, and in Figure 7. A more robust statement of our first year LMC microlensing results is thus that objects in the $2 \times 10^{-4} - 2 \times 10^2 {\rm M}_\odot$ range can contribute no more than $3 \times 10^{11} {\rm M}_\odot$ to the dark halo, where we consider the halo to extend out to 50 kpc. The standard halo has $4.1 \times 10^{11} {\rm M}_\odot$ out to this radius, and so is ruled out as before, but much smaller, all Macho halos, would be allowed. It should be clear that in order to get good information on the Macho fraction of the halo, more work is needed on the total mass of the halo. This requires better measurement of the Milky Way parameters and rotation curve. Microlensing measurements themselves may also be able to help.[23,22,21]

| Model | S | A | B | C | D | E | F | G |
|---|---|---|---|---|---|---|---|---|
| Description | med. | med. | large | small | E6 | max disk | big disk | big disk |
| $\beta$ | - | 0 | -0.2 | 0.2 | 0 | 0 | 0 | 0 |
| $q$ | - | 1 | 1 | 1 | 0.71 | 1 | 1 | 1 |
| $v_a$ (km/s) | - | 200 | 200 | 180 | 200 | 90 | 150 | 180 |
| $R_c$ (kpc) | 5 | 5 | 5 | 5 | 5 | 20 | 25 | 20 |
| $R_0$ (kpc) | 8.5 | 8.5 | 8.5 | 8.5 | 8.5 | 7.0 | 7.9 | 7.9 |
| $\Sigma_0$ (M$_\odot$pc$^{-2}$) | 50 | 50 | 50 | 50 | 50 | 100 | 80 | 80 |
| $R_d$ (kpc) | 3.5 | 3.5 | 3.5 | 3.5 | 3.5 | 3.5 | 3.0 | 3.0 |
| $v_{\rm tot}(R_0)$ (km/s) | 192 | 224 | 233 | 203 | 224 | 234 | 218 | 225 |
| $v_H(50)$ (km/s) | 188 | 199 | 250 | 142 | 199 | 83 | 134 | 167 |
| $v_{\rm tot}(50)$ (km/s) | 198 | 208 | 258 | 155 | 208 | 130 | 160 | 188 |
| $\tau_{\rm LMC}$ ($10^{-7}$) | 4.7 | 5.6 | 8.1 | 3.0 | 6.0 | 0.85 | 1.9 | 3.3 |

**Table 1:** Galactic models for LMC microlensing. Lines 2 - 8 show the model parameters: the asymptotic slope of the rotation curve ($\beta = 0$ flat, $\beta < 0$ rising, $\beta > 0$ falling), the halo flattening ($q = 1$ is spherical), the normalization velocity $v_a$, the halo core radius $R_c$, the solar distance from the galactic center $R_0$, the disk local column density ($\Sigma_0 = 50$ canonical disk, $\Sigma_0 = 100$ extreme maximal disk), and the exponential disk scale length $R_d$. Lines 9 - 12 show useful derived quantities: the total local rotation speed $v_{\rm tot}(R_0) \approx 220$ km/sec, the rotation speed due to only the halo at 50 kpc $v_H(50\,{\rm kpc})$, the total rotation speed at 50 kpc $v_{\rm tot}(50)$, and the predicted microlensing optical depth to the LMC $\tau_{\rm LMC}$.

The limits above are valid whether or not the three events shown in Figure 2 are due to microlensing of halo objects. However, if we make the additional assumption that they are, we can go beyond limits and estimate the Macho contribution to the halo, and also the masses of the Machos. The results obtained, especially on the lens masses, depend strongly on the halo model used, so keep in mind that it is not clear that all three events are microlensing, and it is certainly not known that they are due to objects residing in the galactic halo. Proceeding anyway, we can construct a liklihood function as the product of the Poisson probability of finding 3 events when expecting $N_{exp}$ and the probabilities of drawing the observed $\hat{t}$'s from the calculated model duration distribution.[21,10,11] The resulting liklihood contours can be found in references 10 and 11. The resulting most likely halo fractions, Macho masses, and Macho contributions to the dark halo are shown in Table 2. Our best fit optical depth is also quite model independent. We see that for a standard "S" halo, a macho fraction of $\sim 20\%$ is most likely, with Macho masses in the $0.01 - 0.1 {\rm M}_\odot$ range likely. Note that the errors in these estimates are very large due to the small number statistics, and that there is an enormous additional uncertainty due to the halo model. However, once again, the maximum liklihood estimate of the total mass in Machos is quite model independent, as can be seen from Table 2.

| Model | S | A | B | C | D | E | F | G |
|---|---|---|---|---|---|---|---|---|
| description | med. | med. | large | small | E6 | max disk | big disk | big disk |
| $m_{\rm ML}$ (M$_\odot$) | 0.065 | 0.050 | 0.085 | 0.031 | 0.045 | 0.007 | 0.021 | 0.032 |
| ± | +0.06 −0.03 | +0.05 −0.03 | +0.08 −0.04 | +0.03 −0.02 | +0.03 −0.02 | +0.006 −0.004 | +0.018 −0.010 | +0.03 −0.02 |
| $f$ | 0.19 | 0.16 | 0.12 | 0.31 | 0.15 | 1.1 | 0.50 | 0.29 |
| ± | +0.16 −0.10 | +0.14 −0.08 | +0.10 −0.06 | +0.26 −0.16 | +0.12 −0.08 | +0.82 −0.53 | +0.42 −0.26 | +0.24 −0.15 |
| $v_{\rm ML}(50)$ (km/s) | 82 | 80 | 86 | 79 | 77 | 88 | 95 | 90 |
| ± | +29 −25 | +29 −25 | +31 −27 | +28 −25 | +27 −22 | +28 −24 | +34 −29 | +32 −28 |
| $M_{\rm ML}$ ($10^{10}$M$_\odot$) | 7.6 | 7.4 | 8.5 | 7.2 | 6.8 | 8.9 | 10.0 | 9.2 |
| ± | +6 −4 | +6 −4 | +7 −4 | +6 −4 | +6 −3 | +6 −4 | +9 −5 | +7 −5 |
| $\tau_{\rm ML}/(10^{-8})$ | 8.8 | 9.0 | 9.6 | 9.4 | 9.0 | 9.6 | 9.6 | 9.5 |
| ± | +7 −5 | +8 −5 | +8 −5 | +8 −5 | +7 −5 | +7 −5 | +8 −5 | +8 −5 |

**Table 2:** Maximum likelihood results for the galactic models described in Table 1. The subscript ML indicates the best fit one-dimensional value and the errors are 68% C.L. found by integrating over the orthogonal variable. A Baysean method with the prior $df\,dm/m$ was used. The variables are the best fit Macho mass $m_{\rm ML}$, the best fit halo fraction $f_{\rm ML}$, the best fit rotation speed at 50 kpc due entirely to Machos $v_{\rm ML}(50)$, the best fit "mass" in Machos interior to 50 kpc $M_{\rm ML}(50)$, and the best fit optical depth towards the LMC $\tau_{\rm ML}$.

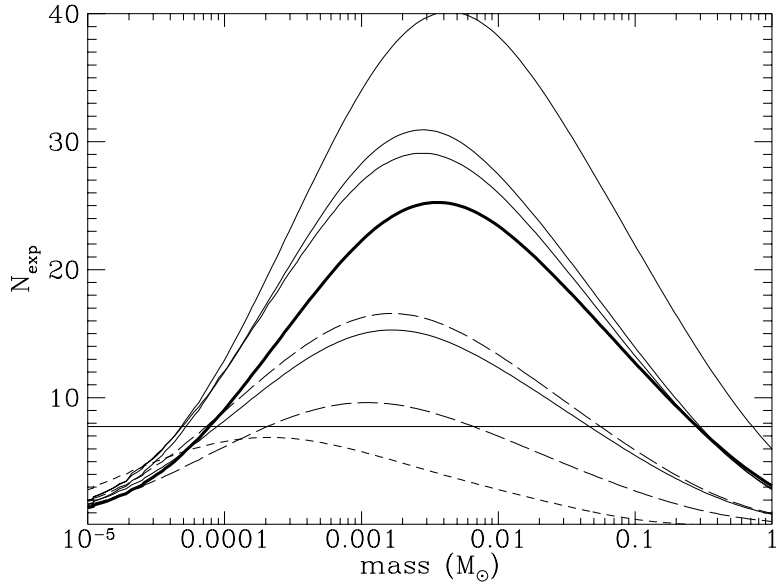

**Figure 6:** Number of microlensing events expected as a function of Macho mass $m$. The thick solid line shows the "standard" model S, the thin solid lines show models B (highest),A,D,C (lowest). The short-dashed line shows the maximal disk model E, and the long-dash lines show the large disk models F,G.

### 7. Interpretation of Bulge Events

The large number of events we (and the OGLE group[7]) have found towards the galactic center came a great surprise to everyone. The line-of-sight toward the bulge goes though the stellar disk, so bulge microlensing is sensitive to halo dark matter, disk stars, and any disk dark matter which might be present. The early predictions[26,27,28] included all these sources, but still predicted many fewer events than have now been observed. It seems the microlensing experiments have discovered a new component of the Milky Way. A standard way of quoting the microlensing probability is the optical depth $\tau$, which is the probability that any given source star is lensed by a magnification of 1.34 or greater. Optical depth has larger statistical errors than the event rate, but has the great advantage of being independent of the masses of the lenses. Early predictions of bulge microlensing were in the $10^{-6}$ range,[26,27,28] while using the sample of events above we find $\tau_{est} = 3.9^{+1.8}_{-1.2} \times 10^{-6}$. We have not finished the complete efficiency calculation for our bulge events, so this estimate uses a sub-sample of 15 giant star events, for which our preliminary efficiencies should be acceptable.[14]

Several models have now been proposed to explain the high microlensing rate. They include[29,30,31]

1) A "heavy" disk. Perhaps the disk of the Milky Way is substantially more massive than normally considered.
2) A "bar" at small inclination. Perhaps the Milky way is not a grand design spi-

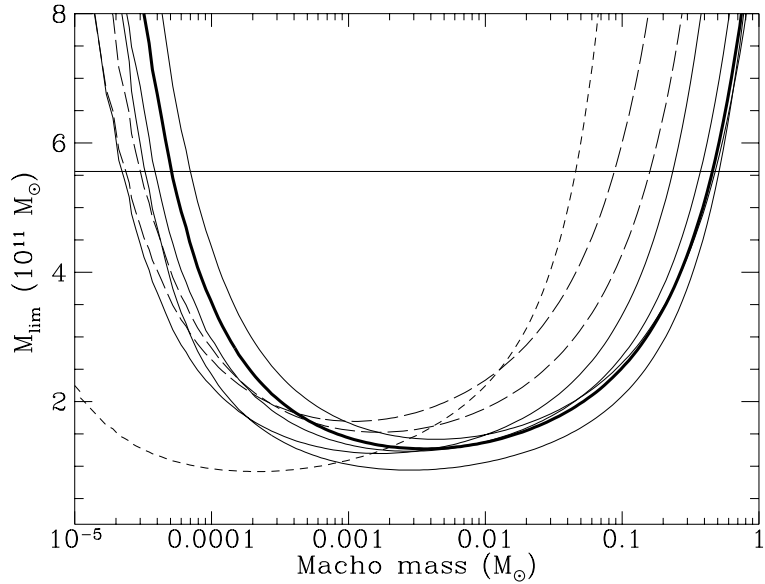

**Figure 7:** 95% C.L. upper limits on the mass in Machos interior to 50 kpc as a function of the Macho mass for the models described in Table 2. The horizontal line at $5.56 \times 10^{11} M_\odot$ shows the mass in a spherical halo with the I.A.U. rotation value of 220 km/sec at 50 kpc. Lines are coded as in Figure 6.

ral as usually assumed, but is a barred spiral, with a very large bar, previously overlooked since it points nearly toward us.

3) A highly flattened, or disk-like halo.

4) Some combination of the above, and/or extra material in the bulge.

Extensive work is being undertaken in trying to resolve these questions. One method is to map out the bulge area with microlensing. A bar-like structure will give a different pattern of microlensing than a disk-like structure. Use of a satellite, or the fine-structure of the microlensing lightcurve have also been suggested.

## 8. Advantages of Having Many Events

There are two main advantages of having several times more events than we originally thought we would have. First, we can do statistical tests on the data. For example, simple geometry predicts a specific distribution of maximum amplifications. Basically, every lens/line-of-sight impact parameter should be equally likely, so the distribution of $u_{min}$'s should be uniform (taking into account that our efficiency for detecting high magnification (low $u_{min}$) events is larger). We have performed a Kolmogorov-Smirnov (K-S) test on the bulge events and find good consistency with the microlensing hypothesis. Thus the microlensing interpretation of these events is greatly strengthened.

The second advantage of having many events, is that rare events can be found. For example events of high magnification or long duration should occur occasion-ally. In Figures 8 and 9 we show unusual events, both of which allow extraction of important information concerning the Macho mass/velocity/distance. In Figure 8 we show an event which lasted about 1/2 year, during which time the Earth had a chance to travel part way around the Sun. This gave our telescope two different perspectives on the lens, resulting in a parallax event. Thus the lightcurve does not fit the naive amplification formula presented earlier. Including the Earth's motion, we find a good fit, and discover that the Macho was moving with a projected transverse velocity of $76 \pm 6$ km/sec. The Macho mass is determined by a combination of this velocity, the event duration, and the distance to the Macho, so for such parallax events there is a one-to-one relationship between the Macho mass and distance. In this case the Macho could be either a brown dwarf star in the galactic bulge, an M-dwarf star at a distance of 2 to 6 kpc, or a more massive star quite nearby.

Another rare type of microlensing event is shown in Figure 9. This lightcurve is characteristic of lightcurves formed by binary lenses. This particular event was also seen by the OGLE group,[7] and detailed analysis will again give information as to the lens masses, distances and velocities. An exciting aspect of such a binary Macho detection, is the possibility of detecting planets around Machos. Given that some of the lenses we observe are in fact low mass stars, it is possible to observe caustic crossing such as illustrated in Figure 9, for planets even down to Earth mass.[32,33] Thus microlensing may well be the best way to discover and get statistics on extra-solar planets.

## 9. Conclusion

The microlensing experiments have given robust and strong limits on the baryonic content of the halo. Much more data from the LMC and SMC will be available soon, so we expect the statistics to improve in the near future. The LMC events, if interpreted as due to halo microlensing, allow a measurement of the baryonic contribution to the halo, which is around 20% for a standard halo. In this case, the most likely Macho contribution to the Milky Way halo mass is about $8 \times 10^{10} M_\odot$, which is roughly the same as the disk contribution to the Milky Way mass. However, the whole story has been made more complicated (and exciting) by the much larger than expected number of bulge microlensing events. These events imply a new component of the Galaxy, and until the nature of this new component is known, unambiguous conclusions concerning the LMC events will not be possible. For example, if the Milky Way disk is much larger than usually considered, a much smaller total halo mass will be required, and so even an all Macho halo might be allowed. Alternatively, the new Galactic component which is giving rise to the bulge events, may also be giving rise to the LMC events, and the Macho content of the halo could be zero. Fortunately, much more data is forthcoming, and many new ideas have been proposed. Microlensing is fast becoming a new probe of Galactic structure, and beside the original potential to discover or limit dark matter, may well produce discoveries such as extra-solar planetary systems.

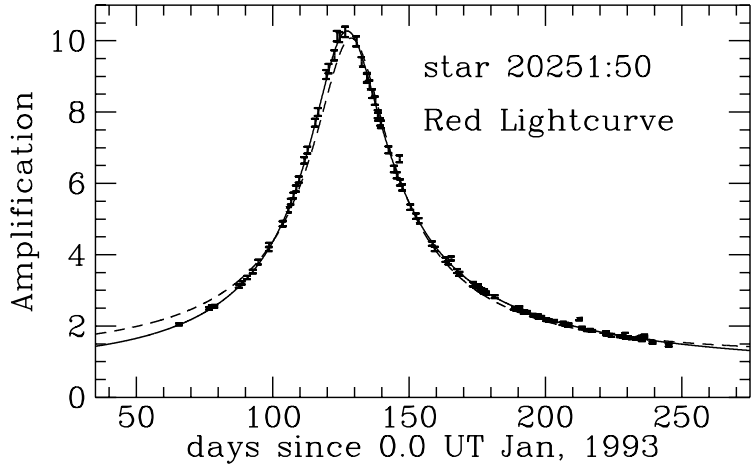

**Figure 8.** The MACHO-Red band light curve for the longest event yet detected by the MACHO project. The solid curve is the fit light curve which includes the effect of the Earth's motion, while the dashed curve is the best fit ignoring the motion of the Earth.

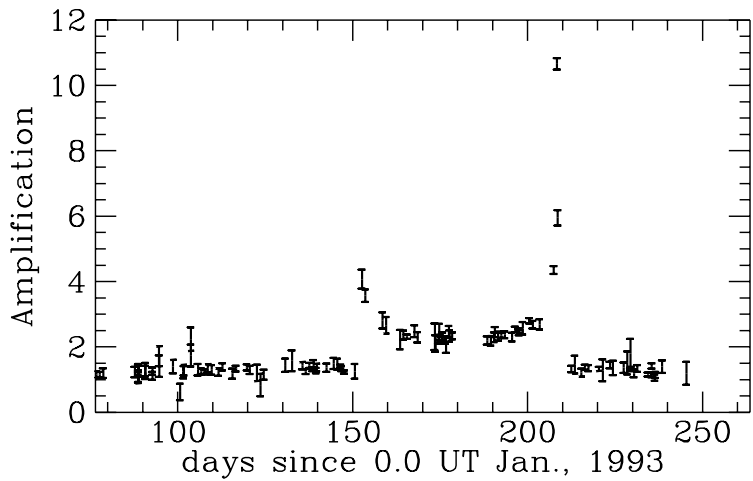

**Figure 9.** The MACHO-Blue band light curve of the binary lens event first seen by OGLE.

## Acknowledgements

Work performed at LLNL is supported by the DoE under contract W7405-ENG-48. Work performed by the Center for Particle Astrophysics on the UC campuses is supported in part by the Office of Science and Technology Centers of NSF under cooperative agreement AST-8809616. Work performed at MSSSO is supported by the Bilateral Science and Technology Program of the Australian Department of Industry, Technology and Regional Development. KG acknowledges DoE OJI, Sloan, and Cottrell Scholar awards.

## References

1. Ashman, K.M., 1992. PASP, 104, 1109.
2. Dekel, A, 1994, ARA&A, 32, 371.
3. Jungman, G. Kamionkowski, M., & Griest, K., 1995, to appear in Physics Reports.
4. for example, Turner, M.S. 1990, Physics Reports 197, 67; Raffelt, G.G. 1990, Physics Reports 198, 1.
5. Paczyński, B, 1986, ApJ, 304, 1.
6. Aubourg, E., *et al.*, 1993, Nature, 365, 623; Beaulieu J.P., *et al.*, 1994, preprint.
7. Udalski, A., *et al.*, 1993, Acta Astronomica, 43, 289; Udalski, A., *et al.*, 1994, Acta Astronomica, 44, 165; Udalski, A., *et al.*, 1994, Acta Astronomica, 44, 227; Udalski, A., *et al.*, 1994, ApJ Lett., in press.
8. C. Alard, private communication.
9. Alcock, C., *et al.*, 1993, Nature, 365, 621.
10. Alcock, *et al.*, 1995, Phys. Rev. Lett. 74, 2867.
11. Alcock, C., *et al.*, 1995, in preparation.
12. Alcock, C., *et al.*, 1995, to appear in ApJ.
13. Bennett, D.P. *et al.*, 1994, Proceedings of the 5th Astrophysics Conference in Maryland: Dark Matter.
14. Alcock, C., *et al.*, 1995, in preparation.
15. Stubbs, *et al.*, 1993, SPIE Proceedings, 1900, 192.
16. Bennett, D.P., 1995, in preparation.
17. Schechter, P.L., Mateo, M., & Saha, A., 1994, PASP, 105, 1342.
18. C.Alcock, 1995 *et al.*, AJ, 109, 1653.
19. Cook, K., *et al.*, 1995, Proceedings of IAU Colloquium 155: Astrophysical Applications of Stellar Pulsation, Cape Town, February 1995, ASP Conference Series, ed. R.Stobie.
20. Griest, K., 1991, ApJ, 366, 412.
21. Alcock, *et al.*, 1995, ApJ 449, 000.
22. Gates, E.I., Gyuk, G. & Turner, M.S., 1995, Phys. Rev. Lett., 74, 3724.
23. Sackett, P. & Gould, A., 1994. ApJ, 419, 648.
24. Evans, N.W., 1993. MNRAS, 260, 191; Evans, N.W., 1994, MNRAS, 267, 333.
25. Evans, N.W., & Jijina, J., 1994. MNRAS, 267, L21.
26. Griest, K., *et al.*, 1991, ApJ Lett., 372, L79.
27. Paczyński, B. 1991, ApJ Lett., 371, L63.
28. Kiraga M., and Paczyński, B. 1994, ApJ Lett., 430, L101.

29. Gould, A., 1994, ApJ Lett., 421, L71; ibid. L75; Gould, A., 1995 ApJ, 441, L21; Gould, A., 1994, preprints.
30. Zhou, H.S., Spergel, D.N. & Rich, R.M., 1994, ApJ Lett., 440, L13.
31. Paczyński, B., *et al.*, 1994 ApJ Lett., 435, L113.
32. Mao, S. & Paczyński, B., 1991, ApJ Lett., 374, L37.
33. Gould, A. & Loeb, A., 1992, ApJ, 396, 101.

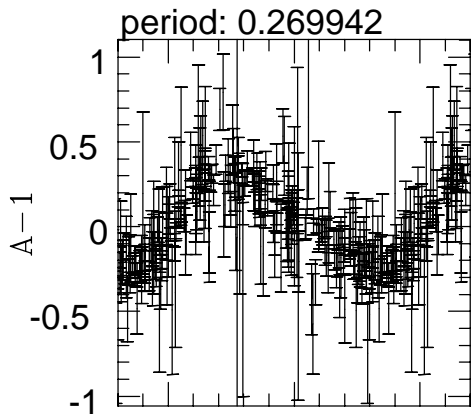
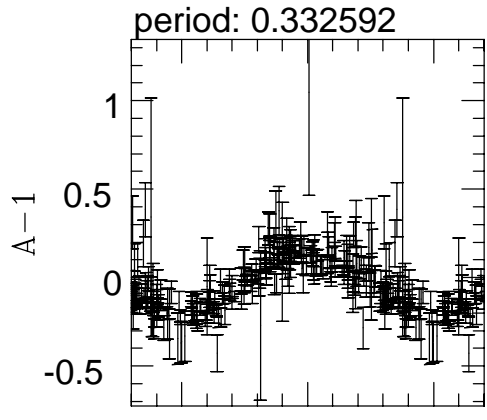
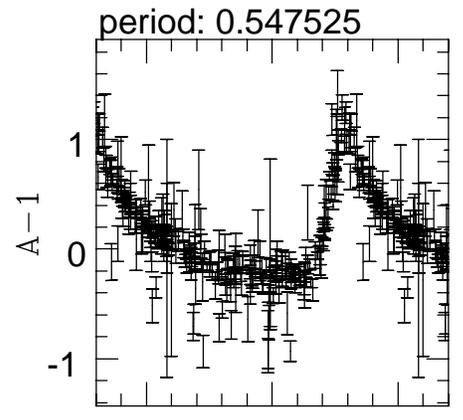
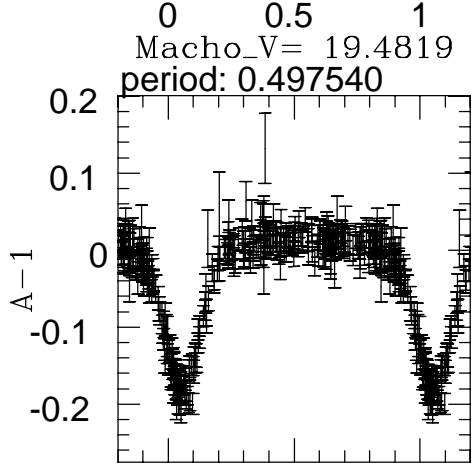
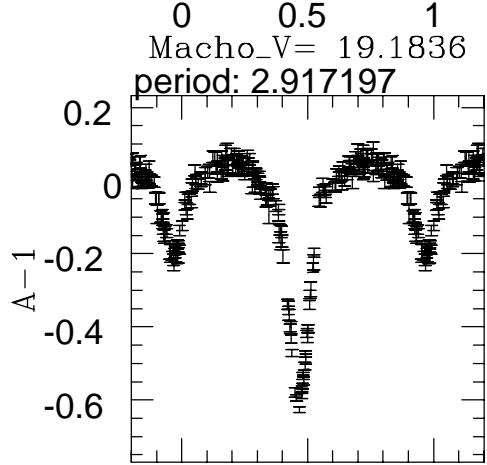
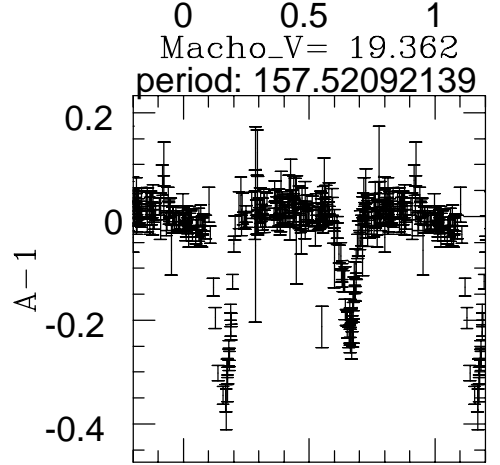
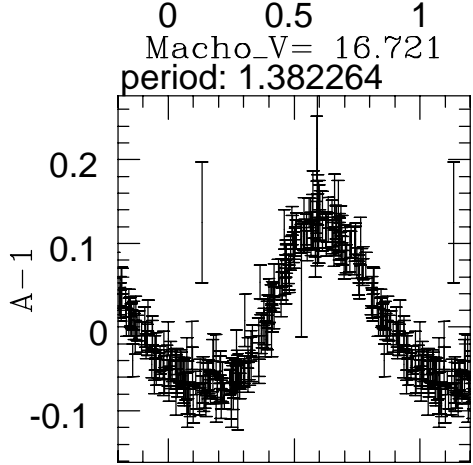
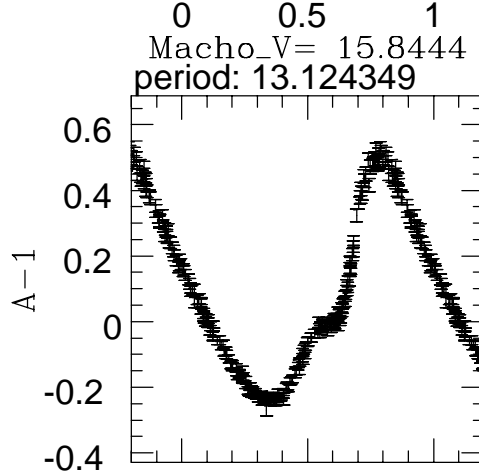
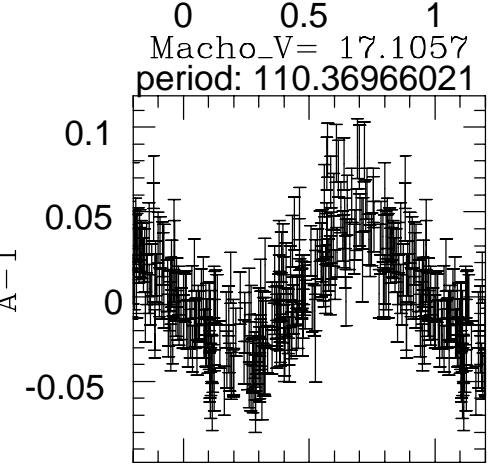

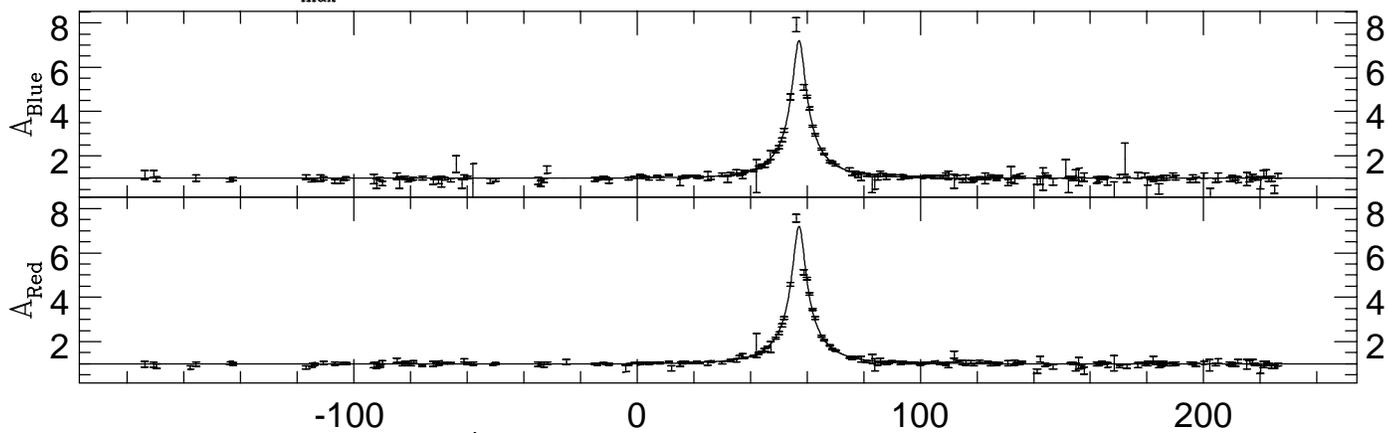
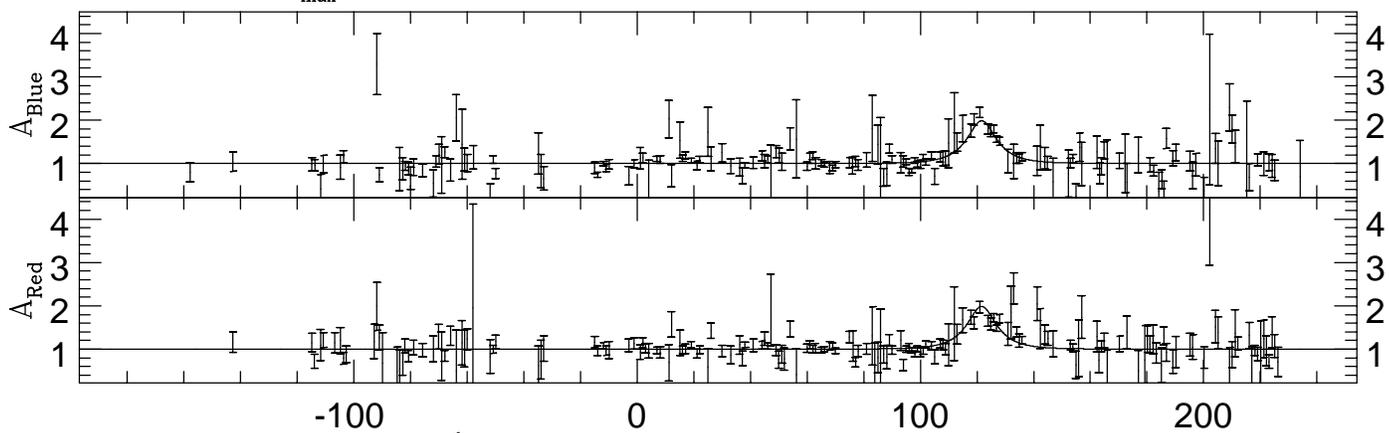
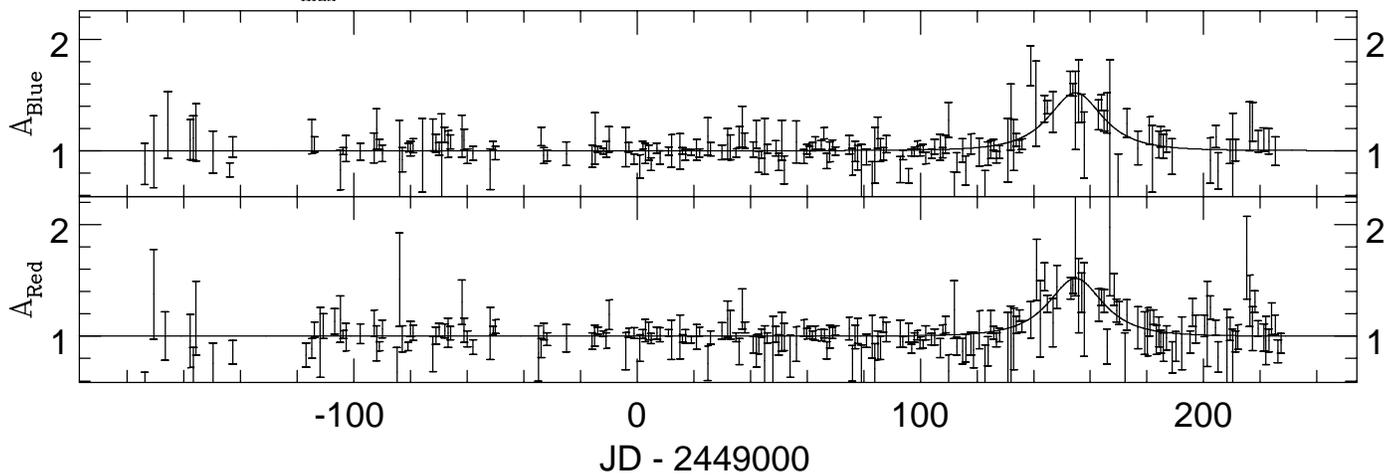

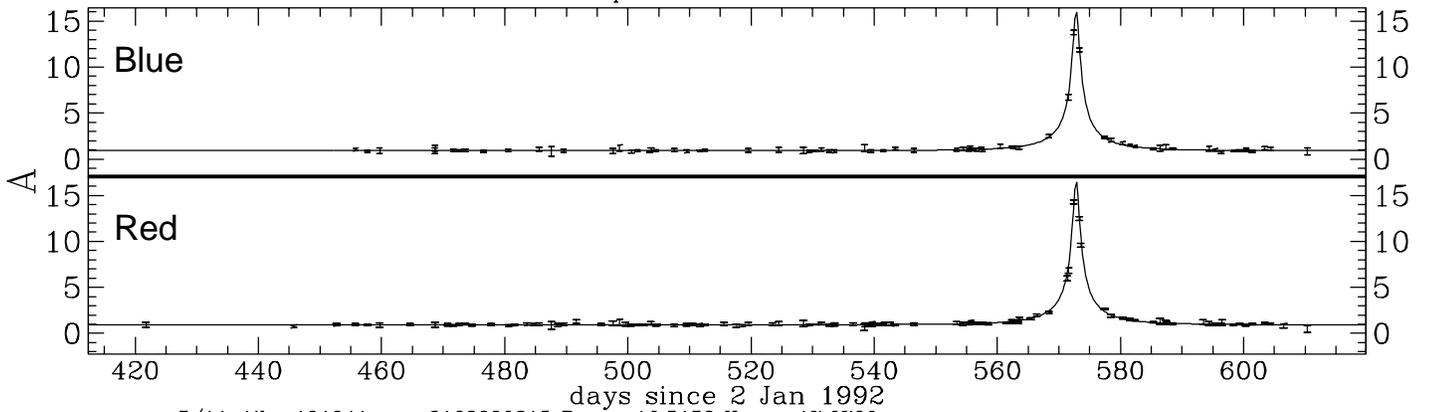

580/21, tile=108054 seq=75910 Rmag=18.8396 Vmag=19.7225
Amax=17.39 that=21.92 chi2dof=0.8887 pchi2dof=1.848 delchi2=1.727e+04

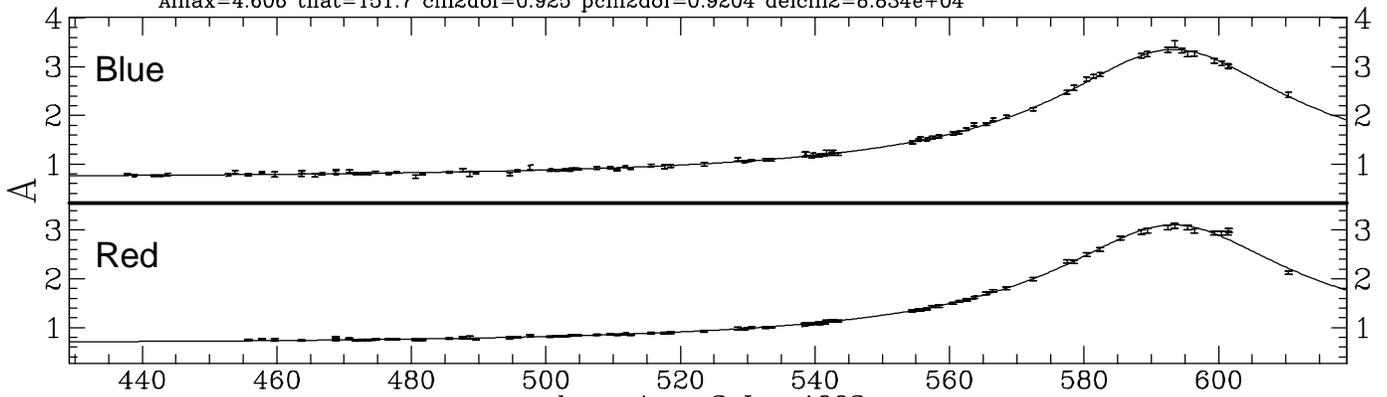

5/11, tile=101041 seq=2168900315 Rmag=16.5152 Vmag=17.6766
Amax=4.606 that=151.7 chi2dof=0.925 pchi2dof=0.9204 delchi2=8.834e+04

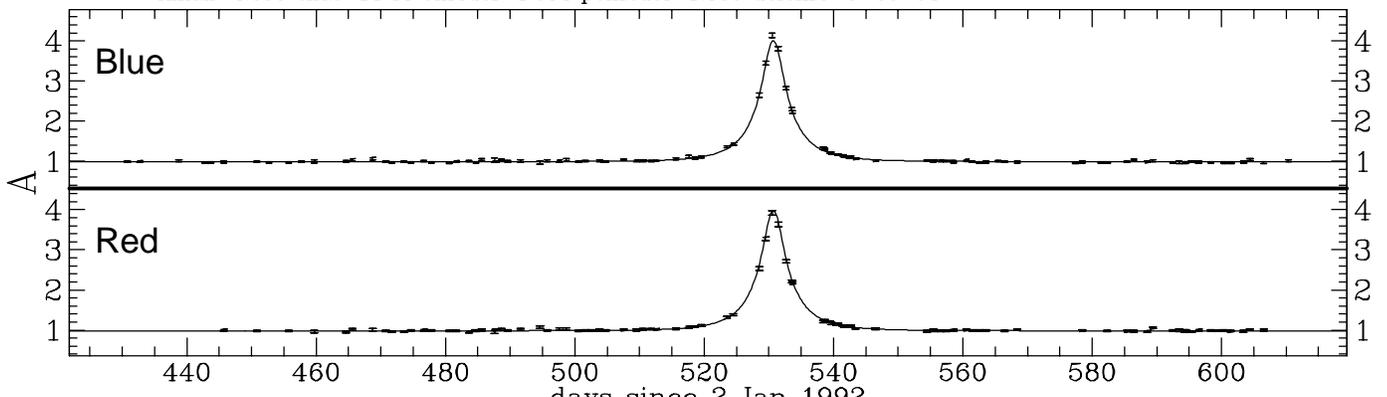

291/21, tile=118018 seq=1827600129 Rmag=17.0592 Vmag=17.6179
Amax=4.055 that=14.44 chi2dof=1.034 pchi2dof=1.336 delchi2=2.73e+04

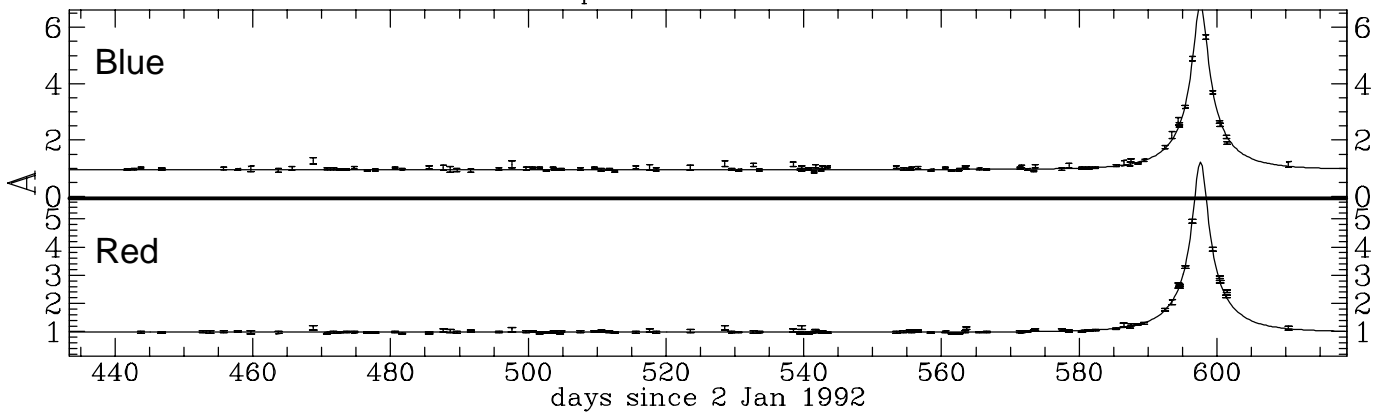

731/21, tile=128027 seq=2166600982 Rmag=17.998 Vmag=18.8313
Amax=7.115 that=16.38 chi2dof=1.686 pchi2dof=2.909 delchi2=3.158e+04